\documentclass[preprint]{aastex}

\usepackage[chatter]{rotating}
\usepackage{amssymb,amsmath}
\usepackage{graphicx}
\usepackage{natbib}
\usepackage{array}

\begin{document}


\title{EUV Observational consequences of the spatial localisation of nanoflare heating within a multi-stranded atmospheric loop}


\author{Aveek Sarkar and Robert W Walsh}
\affil{Centre for Astrophysics,
University of Central Lancashire,  
 Preston, Lancashire, PR1 2HE, UK}
\email{asarkar1@uclan.ac.uk}



\begin{abstract}
Determining the preferred spatial location of the energy input to solar coronal loops would be an important step forward towards a more complete understanding of the coronal heating problem. Following on from \citet{sw08} this paper presents a short ($10^9$cm$\equiv 10$Mm) ``global loop'' as $125$ individual strands, where each strand is modelled independently by a one-dimensional hydrodynamic simulation. The strands undergo small-scale episodic heating and are coupled together through the frequency distribution of the total energy input to the loop which follows a power law distribution with index $\sim 2.29$. The spatial preference of the swarm of heating events from apex to footpoint is investigated. From a theoretical perspective, the resulting emission measure weighted temperature profiles along these two extreme cases does demonstrate a possible observable difference. Subsequently, the simulated output is folded through the TRACE instrument response functions and a re-derivation of the temperature using different filter-ratio techniques is performed. Given the multi-thermal scenario created by this many strand loop model, a broad differential emission measure results; the subsequent double and triple filter ratios are very similar to those obtained from observations. However, any potential observational signature to differentiate between apex and footpoint dominant heating is possibly below instrumental thresholds. The consequences of using a broadband instrument like TRACE and Hinode-XRT in this way are discussed.
\end{abstract}





\section{Introduction}

Considerable effort has been devoted to date to determine the nature of solar magnetic plasma loops, the building blocks of the solar atmosphere. Particular interest has centered upon the examining of any temperature variation along an observed loop which could be related, through model comparison, to a possible prescribed heat input that has a well defined spatial localisation associated with it. Thus, there are two important avenues of investigation linked together here. Firstly, one can take an observed thermal profile say and endeavour to reproduce the temperature structure via a hydrostatic model. \citet{priestna,priestap} outlines evidence that the heat input in their large ($>700$Mm) Yohkoh soft-Xray loop was distributed uniformly in the upper coronal portion ($1.6-2.2$ MK). This has been challenged by \citet{markus01} and \citet{reale02}, who argued for footpoint and apex dominant heating respectively from the same dataset. This has been further investigated by \citet{Soteris08} who, using Bayesian analysis methods, also argues for footpoint heating. Of particular relevance to the work outlined in this paper are the results of \citet{winebarger04}, who studied the dynamic behaviour of the  temperature and density of an impulsively heated 1D hydrodynamic loop. From their analysis they  conclude that TRACE observations of cooling loops are unable to provide adequate information about the magnitude, duration and location of the earlier energy release. Also, \citet{parenti06} employ forward modeling techniques upon a nanoflare-heated multistranded loop to suggest that only high temperature coronal lines are capable of ``remembering'' the original heating signature.

Secondly, one can take a theoretical model and fold the resulting plasma parameters through the instrument response functions to mimic what a given telescope would observe. With that in mind, it is very important to be aware of what current observational results are showing us. Consider concentrating upon the interpretation of EUV imaging data only. A number of authors (\citet{markus05}, \citet{jane07}, \citet{jane08}, \citet{schmelz07}) have used TRACE filter ratio data in an attempt to investigate the physical properties of coronal loops with mixed success. Whether one uses double filter (e.g. \citet{schmelz07}) or triple filter (e.g. \citet{jane07}, \citet{jane08}) ratios, it appears to be the case that a consistent, unique temperature determination is not possible. On top of that, another important factor is whether a loop is a single, monolithic structure or is the amalgamation of many sub-resolution evolving strands. It has been observed that the cooling time for an EUV loop is longer than the cooling time for a monolithic plasma structure (e.g. \citet{winebarger03a}, \citet{ignacio09}), hence leading to the suggestion that this extended cooling period is the result of a bundle of successive, locally impulsively heated (and then cooling) plasma strands (\citet{warren02}).

The concept of a loop as consisting of numerous sub-resolution elements was investigated initially by \citet{cargill94} and \citet{CK07}, where the authors assumed each strand could be represented by a single temperature and density value only. By heating each strand with several nanoflare-like energy events and allowing them to cool down by conduction and radiation, these authors constructed and investigated the behaviour of a multi-stranded global loop. Subsequently, \citet{spiros05} used their nanoflare heated multistranded model to show that the TRACE and \textit{Yokoh} SXT emission was only weakly affected by the various dominant heat deposition locations. \citet{warren06} also used the nanoflare heated multi-stranded model to show flare lightcurve can be better explained when the loop is divided into subresolved strands and heated impulsively. 

In this paper, we consider once again the multistrand loop model introduced in \citet{sw08} (heareafter SW08) but here investigate the physical and observable consequences of spatially localising the nanoflare-heat input to either the loop structure apex or footpoint areas. Section 2 outlines briefly the multistrand model while in Section 3, the results for apex and footpoint dominant heating are outlined. The observed consequences of such modelling are described in Section 4, concentrating upon recreating the filter ratio methods from \citet{jane07} and \citet{reale07}, while further work in this area is outlined in Section 5.

\section{Numerical simulation of multi-strand loop}

\subsection{Model description}

The particular nanoflare multi-strand model employed in this paper was developed and outlined in SW08. Here we describe briefly the important aspects. The model simulates a short loop of length $10$ Mm. This loop consists of $125$ strands where each individual element evolves hydrodynamically independently from each other; that is, there are $125$ separate one dimensional hydrodynamic simulations. However the strands are coupled together through the frequency distribution of the total energy input across the loops. Our 1D hydrodynamical code is based on the Lagrange-Remap method (\citet{toni}), which is able to capture the shock fronts developed during the episodic heating events. 

At the beginning of every simulation all strands are at $10^4$ K and the plasma velocity is zero. The pressure and density maintains an exponential profile to represent gravitationally stratified plasma. There exists a highly dense plasma region ($\sim 1Mm$ long) at each end to mimic the chromosphere in the usual way. 

\subsection{Heat input to the strands}

It is widely accepted that the often subtle interactions between the solar atmospheric magnetic field and the photospheric motions into which it is embedded, play the vital roles in determining a potential coronal heating mechanism. The question arises as to which of the competing physical timescales (driving versus alf\'{v}en) is dominant and hence whether AC (wave heating; \citet{muller94}) or, DC (magnetic reconnection of a tangled field or nanoflares; \citet{parker83,parker88}) heating is taking place. In this paper we concentrate upon the DC mechanism of small, localised energy bursts occurring along a loop structure.

\begin{figure}
\centering
\includegraphics[width=0.65\textwidth,angle=90]{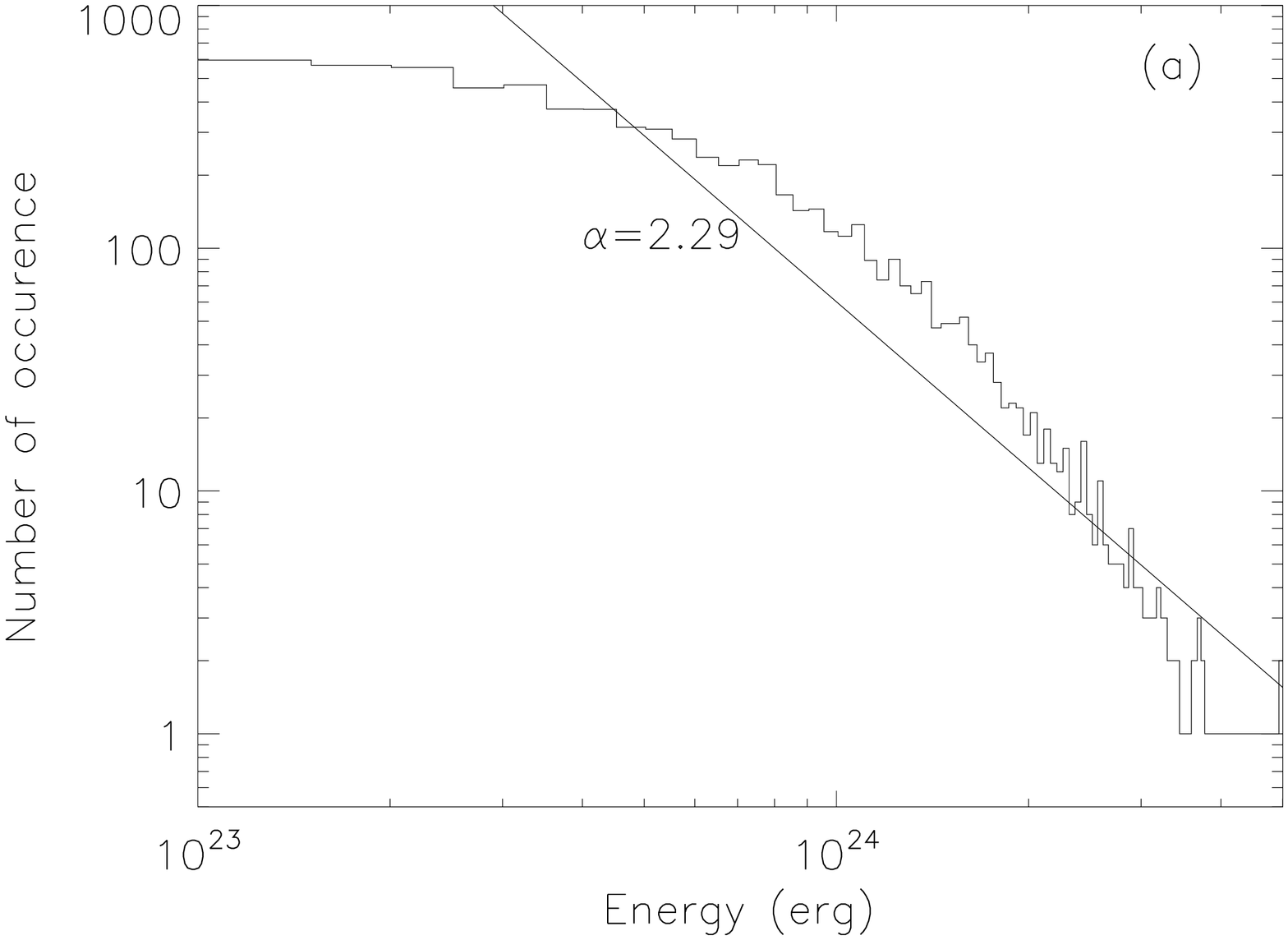}
\includegraphics[width=0.6\textwidth,angle=90]{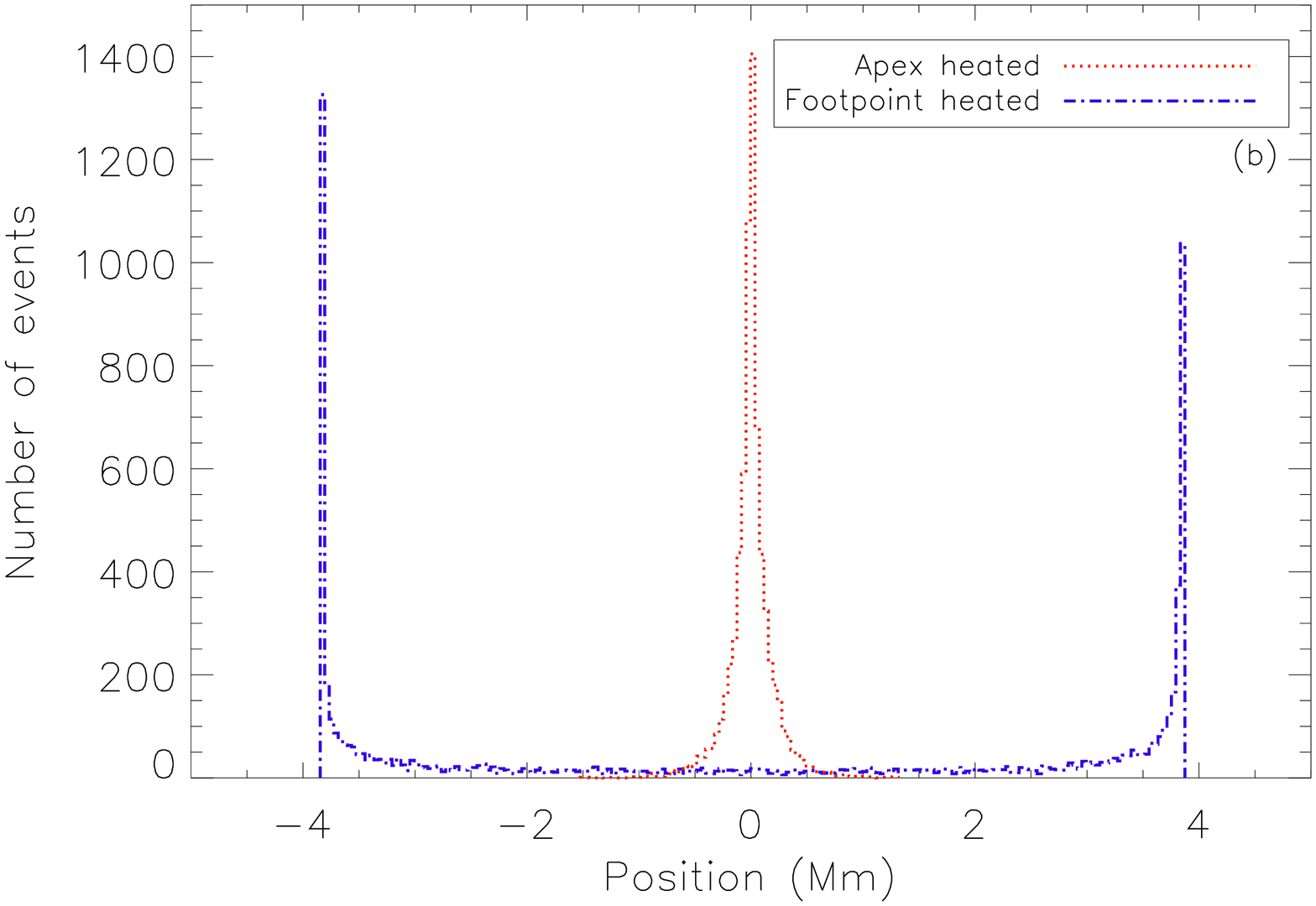}
\caption{(a)Energy histogram for the size of a heating event versus the number of times it occurs during the simulation across all 125 strands. This has been fitted with a straight line, to show the power law slope of $\alpha=2.29$; (b)Location histogram displaying the preferential location of the nanoflare events within the loop. Two extremes are considered -apex (......) and footpoint (\_.\_.\_.\_) dominant heating. \label{fig1}}
\end{figure}

Hudson (1991) suggests that in order for nanoflares to provide the dominant heat input to the corona, the power law index associated with the frequency distribution of the energy event size should be more than $2$. In SW08, we studied the effect of varying this index from $0$ to $3.29$. In the following simulations we keep $\alpha=2.29$ (Case B in SW08) with a nanoflare energy range of $10^{23}$ to $5 \times 10^{24}$ erg. Figure \ref{fig1}(a) displays a histogram of event size versus occurrence rate for the simulations. In SW08, the event locations were spread randomly along the strands between $-3.9$ to $3.9$Mm- this produces an effective spatially uniform heat input to the loop. In these investigations, we now localise spatially where the events can occur. 

We consider two extreme cases as shown in Figure \ref{fig1}(b) which is a ``location" histogram. Firstly, there is the ``footpoint dominant" heating (FDH) case where the vast majority of the nanoflares occur within a narrow region of $\pm 3.5$ to $\pm 3.9$Mm. Secondly, there is the case where the episodic energy events are clustered close to the loop apex at $s=0$ (termed ``apex dominant" heating or ADH).  Note that \textit{(i)} in all cases, the overall total amount of heat provided to the strands is the same- the main investigation here is the impact on what we could observe when the  heat input is distributed differently; \textit{(ii)} the nanoflare lifetimes are chosen randomly over a range of $50$ to $150$ s and release their energy over a length scale of $0.02$ Mm. The consequences of this spatial variation in the heat input are discussed in the following sections. 

\section{Results of multi-strand model} 

\subsection{Effect on loop apex temperature}

\begin{figure}
\centering
\includegraphics[width=0.42\textwidth,angle=90]{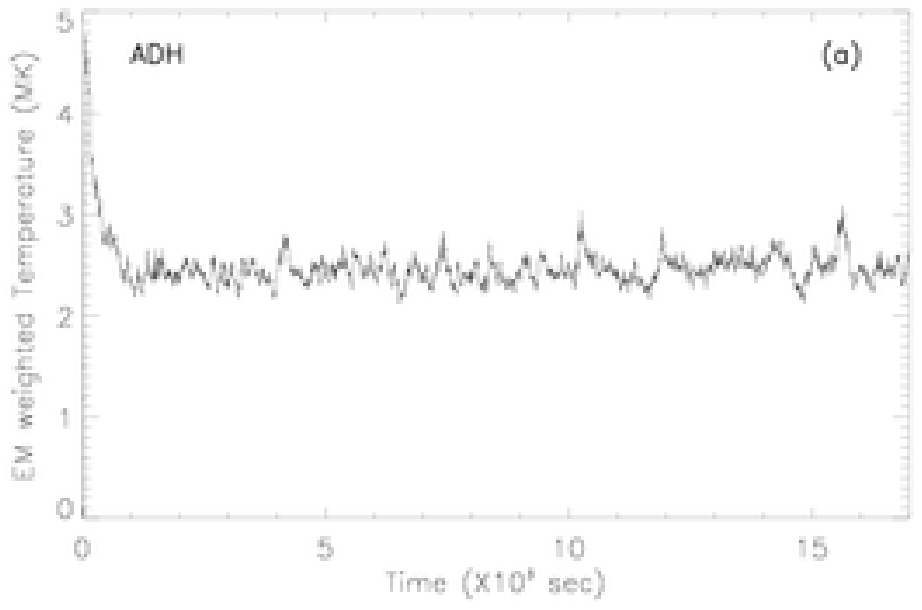}
\includegraphics[width=0.42\textwidth,angle=90]{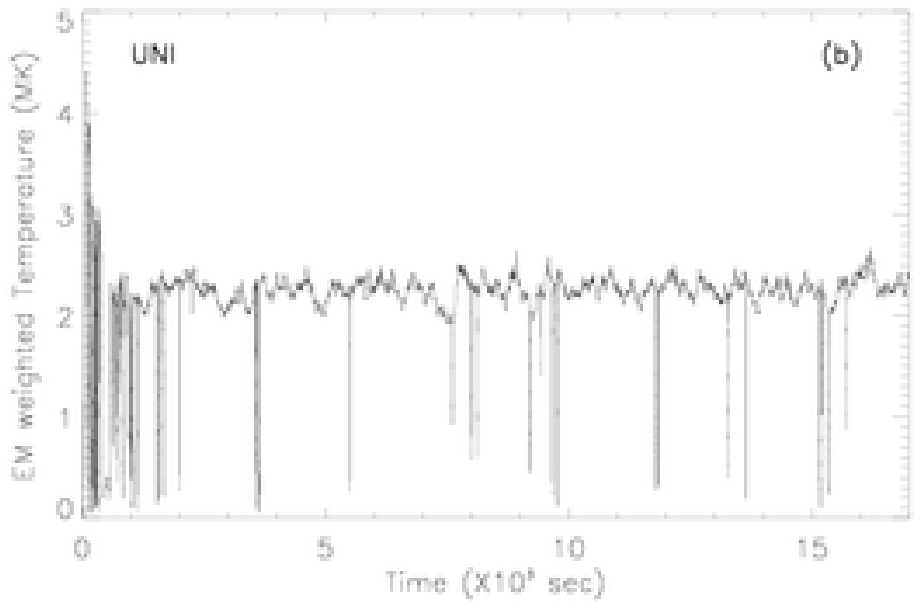}
\includegraphics[width=0.42\textwidth,angle=90]{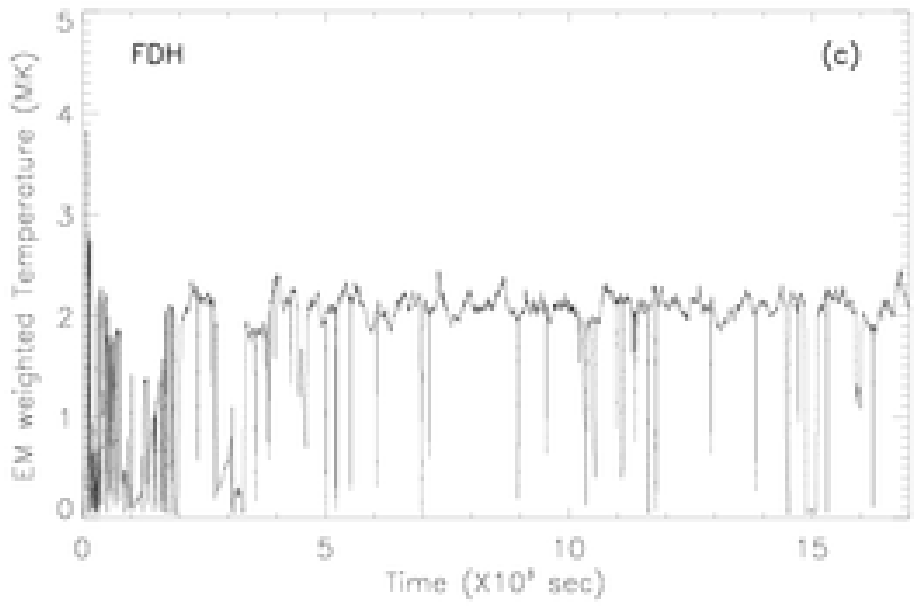}
\caption{Time evolution of $\overline{T}_{EM}$ at the loop apex for three cases (a) apex dominant heating (ADH) (b) spatially uniform heating (UNI) and a (c) footpoint dominant heating (FDH). \label{fig2}}
\end{figure}

As a reminder, SW08 defines the emission measure weighted temperature ($\overline{T}_{EM}$) as

\begin{equation}
\overline{T}_{EM}=\dfrac{\displaystyle \sum_{i=1}^{125} n_{i}^2(s,t)dl(s)T_i(s,t)}{\displaystyle \sum_{i=1}^{125} n_{i}^2(s,t)dl(s)}
\end{equation}

where $n$ indicates particle density, $dl$ is the physical length of the computational grid, $T$ is the temperature, $s$ the position along the loop, $t$ is the time and $i$ is the number of the strands. Subsequently, Figure \ref{fig2} plots $\overline{T}_{EM}$ at the loop apex for three distinct cases. Note that Figure \ref{fig2}(b) displays the same plot as Figure 6 (top left) in SW08 which is the case of spatially uniform heating (UNI). Of specific interest in SW08 the sudden dips in temperature, shown to be related directly to cool plasma blobs travelling along individual strands. It was demonstrated that these are generated when nanoflares release their energy further down the legs of the strands towards the chromospheric region. This in turn causes chromospherically evaporated blobs which travel along the strand and hence effect the calculation of $\overline{T}_{EM}$. 
 
 Figure \ref{fig2}(a) shows the ADH case. There are two aspects to note. Firstly given the spatial preference for the nanoflare bursts to occur far from the loop legs, no plasma blobs are created along any of the $125$ strands. Secondly, there is an increase in the apex $\overline{T}_{EM}$ up to $\sim 2.4$MK compared to $\sim 2.2$MK in the uniform heating case. In contrast, Figure \ref{fig2}(c) displays the FDH case. Compared to the UNI case, there are more plasma blob events influencing $\overline{T}_{EM}$as should be the case given the preferred energy deposition locations. Also, the average $\overline{T}_{EM}$ is depressed further to $\sim 2.1$MK. Examining the resulting $\overline{T}_{EM}$ differences further, Figure \ref{fig3} plots the time-averaged $\overline{T}_{EM}$ profile along half of the loop length [-5, 0]Mm for ADH and FDH over a period of $\sim 1500$s from a start time of $7.3 \times 10^3$s- this time window is chosen to avoid any major plasmoidal flows along the strands. We notice that there is a distinct difference in these thermal profiles. As mentioned by other authors (\citet{priestap},\citet{robert99}), the preferred location of the dominant heat input to the strands (and subsequently to the amalgamated loop) does have a measurable effect on the averaged resulting temperature structure.
   
   For the ADH case, the nanoflare heating is taking place in an environment where thermal conduction dominates as the energy mechanism. Subsequently, the average temperature and the temperature gradient will increase to ``remove" the deposited heat. In contrast, for the FDH case, the thermal profile is much flatter in the ``coronal part" of the loop. The nanoflare energy deposition is taking place close to the strand/ loop legs where radiation will play a more important role as well as a limited conductive loss.

\section{EUV imaging observations and the rederivation of the loop temperature from filter-ratio method}
In recent years, it has become possible to undertake forward modelling of simulation data by folding a calculated temperature/density through the response function for our instrument of choice. Thus, let us consider producing the synthetic emission for three passbands on the Transition Region and Coronal Explorer (TRACE) mission- 171, 195 and 284 $\mathring{A}$. Initially
\begin{equation}
 I_{\lambda,i}=G_{\lambda}(T) n_{i}^2(s,t)ds
\end{equation}
where $I_{\lambda,i}$ is the intensity of the strand $i$ at passband $\lambda$. $G_{\lambda}(T)$ is the temperature response function at passband $\lambda$, $n_i$ is the density of the strand $i$, which is a function of space($s$) and time($t$) and $ds$ is the computational grid length. While calculating the intensity, Feldman (1992) coronal abundances and Arnaud \& Raymond (1992) ionization balance are considered. After calculating the emission from all the strands, one can derive the overall intensity of the global loop as,

\begin{equation}
I_{\lambda}=\sum_{i=1}^{125}I_{\lambda,i}
\end{equation}

Once the loop intensity for each passband is derived, we can employ different ratio techniques to see the effect of a multistranded model. However it must be kept in mind that one of the main assumptions behind all of the following ratio methods is the notion that the volume of plasma under investigation in an instrumental pixel is isothermal. This is not the case for our multi-strand, multi-thermal model. However, it is instructive to examine what the subsequent consequences of this in relation to how current observational datasets are being analysed.

\subsection{Single filter ratio}
The single filter ratio method has been used by several authors (e.g. \citet{markus00}) to estimate the temperature of a target loop from EUV imaging data. Using the synthesised EUV emission produced from our model as outlined above, Figure \ref{fig4} displays the time evolution of three filter ratios for the (a) ADH and (b) FDH cases for a TRACE-sized pixel (1 arcsec) placed at the loop apex. A time integration (exposure time) of $60$s has been employed.

There are several aspects to note. Firstly, compared to the $\overline{T}_{EM}$, none of the three filter ratios provided a ``correct" estimate of the loop temperature. Whereas the average $\overline{T}_{EM}$ for the ADH (FDH) case was $\sim 2.4$MK ($2.0$MK), the three filter ratios vary around $\sim 1.8$MK (284/195), $\sim 1.6$MK (284/171) and $\sim 1.25$MK (195/171) respectively. Observationally it is always seen (e.g. \citep{schmelz03,schmelz07a}) from TRACE and EIT active region data that the temperature obtained from the 195 to 171 ratio is always statistically indistinguishable from $1.2$MK and the 195 to 284 ratio always biased towards $1.8$MK.  There appears to be no distinct observational difference in the filter ratio temperature between ADH and FDH, apart from the fact that the ADH variation appears to be larger, a natural consequence as the dynamic heating is taking place directly at the apex.

Now consider deriving the thermal profile along the loop from these synthesised TRACE emissions. Employing the $284/195$ $\mathring{A}$ ratio with a $60$s integration time, Figure \ref{fig5} displays the resulting temperature variation along half the loop and slightly above the chromospheric boundary. Comparing this to Figure \ref{fig3}, the resulting thermal profiles for ADH and FDH are different. However, the temperature discrimination of TRACE would not be adequate to detect this slight difference of $\sim 0.1$ MK, which would be of the order of the observed error bar.

\subsection{Color-Color method}

Given the apparent problems with the single filter ratio, a double filter ratio method has been employed that combines all three ratios (\citet{chae02}, \citet{jane07}). Here single filter ratios are plotted against one another to create a unique temperature-dependent color-color (C-C) curve. In this case, we define  
\begin{equation}
C_1=\dfrac{G_{284}}{G_{195}}=\dfrac{I_{284}}{I_{195}}
\end{equation}
\begin{equation}
C_2=\dfrac{G_{195}}{G_{171}}=\dfrac{I_{195}}{I_{171}}
\end{equation}
where $C_1$ and $C_2$ are now functions of temperature. 
Figure \ref{fig6} plots the theoretical $C_1/C_2$ curve with the corresponding values derived from (a) ADH and (b) FDH scenarios, where the same TRACE pixel-sized portion of the loop apex is used over an integration time of $\sim 1500$s. In this case, a much longer integration time of $\sim 1500$s is used to include any longer term variability.

In both cases, the synthesised data points cluster in a region far from the C-C curve. As the bounding box shows, there is a slight difference between ADH and FDH, corresponding to a small decrease in both ratios. This is due to the perceived small decrease in ``temperature'' indicated in Figure \ref{fig5}.

It should be noted that this type of clustering behaviour has been observed in TRACE observational data. For example, in Figure 5 of \citet{jane07}, there is a clear demonstration of clustering of datapoints along loop structures observed at the solar limb.

\subsection{Combined filter ratio method}

Given the above unsuccessful attempts to predict a temperature from single and double filter ratio methods, it is instructive to consider a combined filter ratio (CFR) as introduced by \citet{reale07}. 

It must be noted that those authors employ this method upon XRT observations where they have many more filters available compared to the three TRACE passbands considered here. Briefly, CFR is the ratio between the geometric mean of the emission detected in all available filters and the emission $I_{\lambda}$ in a single ($\lambda$-th) filter given by

\begin{equation}
\label{eq:cfr}
CFR_{\lambda}(T)=\dfrac{(\displaystyle \prod_{\lambda =1}^{n}I_{\lambda})^{1/n}}{I_{\lambda}}=\dfrac{(\displaystyle \prod_{\lambda =1}^nG_{\lambda}(T))^{1/n}}{G_{\lambda}(T)}
\end{equation}

Thus in this case, $n=3$. A further improvement is obtained by taking the product of the ratios computed for the two filters. The resulting combined improved filter ratio is defined as 
\begin{equation}
\label{eq:cifr}
 CIFR(T)=CFR_{1}(T) \times CFR_{2}(T)
\end{equation}
Figure \ref{fig7} plots the CIFR for the FDH only- there is no significant difference between these curves and the ADH case. Note that while plotting we again consider the average value over a TRACE-sized pixel placed at the loop apex ($s=0$) location. The fluctuations observed during the first $1000$s are due to the high fluctuations in density at the beginning of the simulation. CIFR predicts values for the temperature that are quite different than the values predicted by the corresponding filter-ratio temperature (Figure \ref{fig4}(b)).  In this case, the calculated temperatures are about $1.65$MK, $1.5$MK and $2.1$MK for CIFR($171 \times 195$), CIFR($195 \times 284$) and CIFR($284 \times 171$) respectively. The fluctuations in the CIFR temperature evolution are less than the single filter-ratio as well.

\section{Summary and discussion}

\begin{figure}
\centering
\includegraphics[width=0.85\textwidth,angle=90]{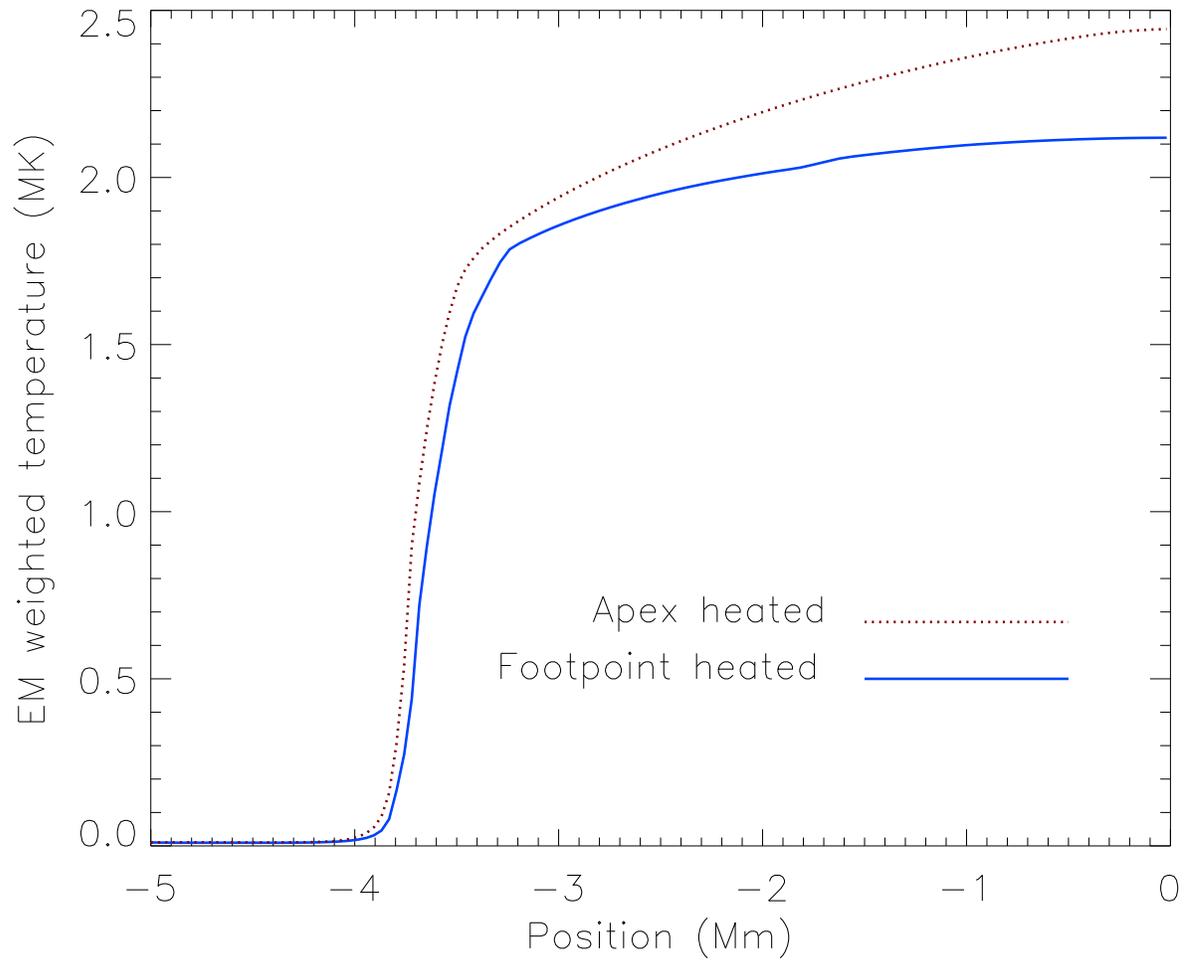}
\caption{ Time averaged emission measure weighted temperature profile along the half loop length ADH (.....) and FDH (\_\_\_\_) over a time period of $\sim 1500$s. \label{fig3}}
\end{figure}

In this paper we have examined a multi-stranded loop model where the episodic heat input to the structure is deposited at two extreme locations; namely, at the strand/loop apex (ADH) and footpoint (FDH). The time averaged EM weighted temperature along the loop that results from the overall effect of all $125$ strands was calculated. As shown in Figure \ref{fig3}, the ADH thermal profile has a steeper temperature gradient in the coronal ($>10^6$ K) part of the loop than that resulting from FDH. This mirrors similar results from hydrostatic modelling of spatially dependent heating in loops (\citet{robert99}, \citet{markus01}).

Subsequently, the multi-strand simulation values were folded through the instrument response function for three EUV TRACE passbands ($171$, $195$, $284$ $\mathring{A}$) and the cumulative intensity across all strands calculated. The resulting emission was then used to calculate a loop plasma temperature by three filter ratio techniques- single, double and combined improved. The resulting values and evolution of a TRACE-sized pixel at the loop apex is very representative of what we evaluate from TRACE loop observations.

Also it is shown that, using the single filter ratio, the derived temperature along the length of the loop for FDH and ADH will not be differentiable observationally. This conclusion is similar to that in \citet{spiros05} who, using their loop model showed that TRACE \& Yohkoh-SXT observation of loops depend only weakly upon the spatial distribution of nanoflare heating. Also \citet{winebarger04} and \citet{parenti06} point out the TRACE and low temperature spectroscopic observations (where radiation is the dominant cooling phenomenon) are unable to retain information about the nature and location of the original heat input.

\begin{figure}
\centering
\includegraphics[width=0.65\textwidth,angle=90]{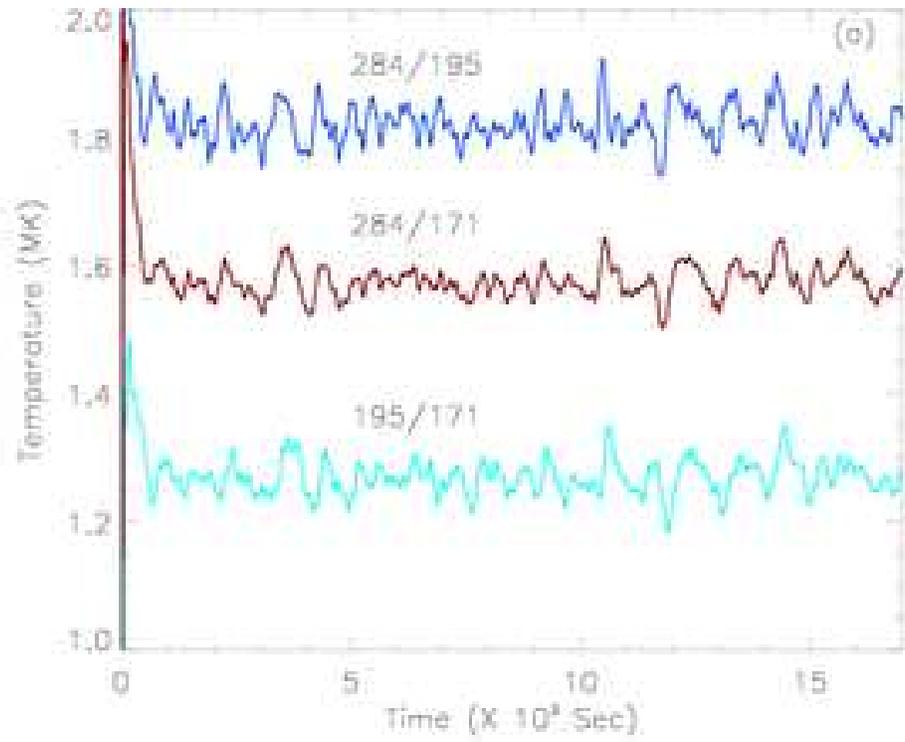}
\includegraphics[width=0.65\textwidth,angle=90]{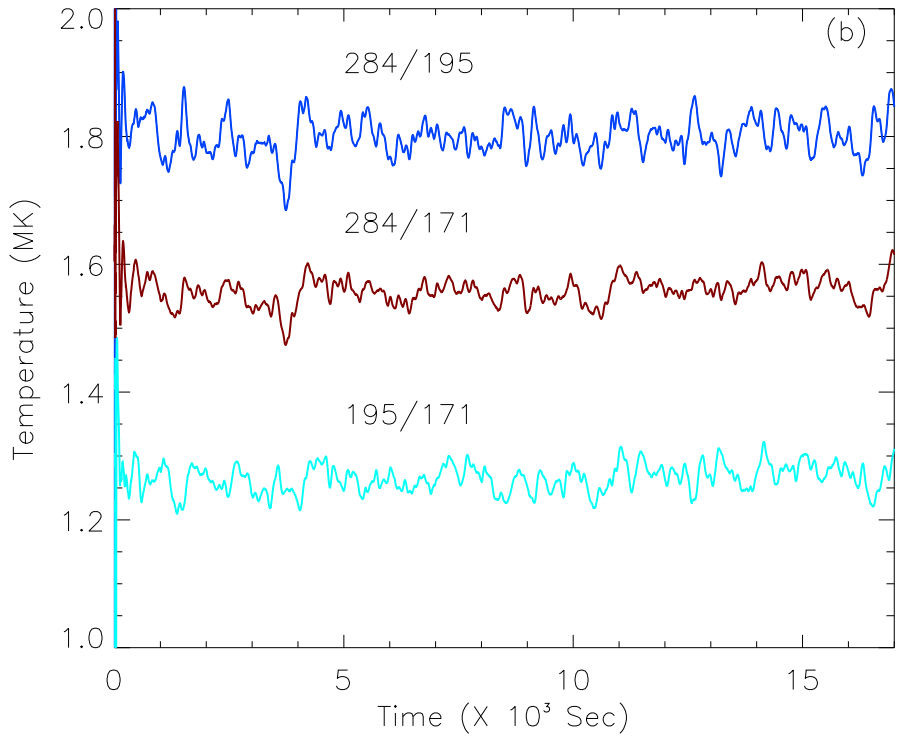}

\caption{Single filter ratio temperature evolution at the loop apex for (a) ADH and (b) FDH loop. \label{fig4}}
\end{figure}

\begin{figure}
\centering
\includegraphics[width=0.85\textwidth,angle=90]{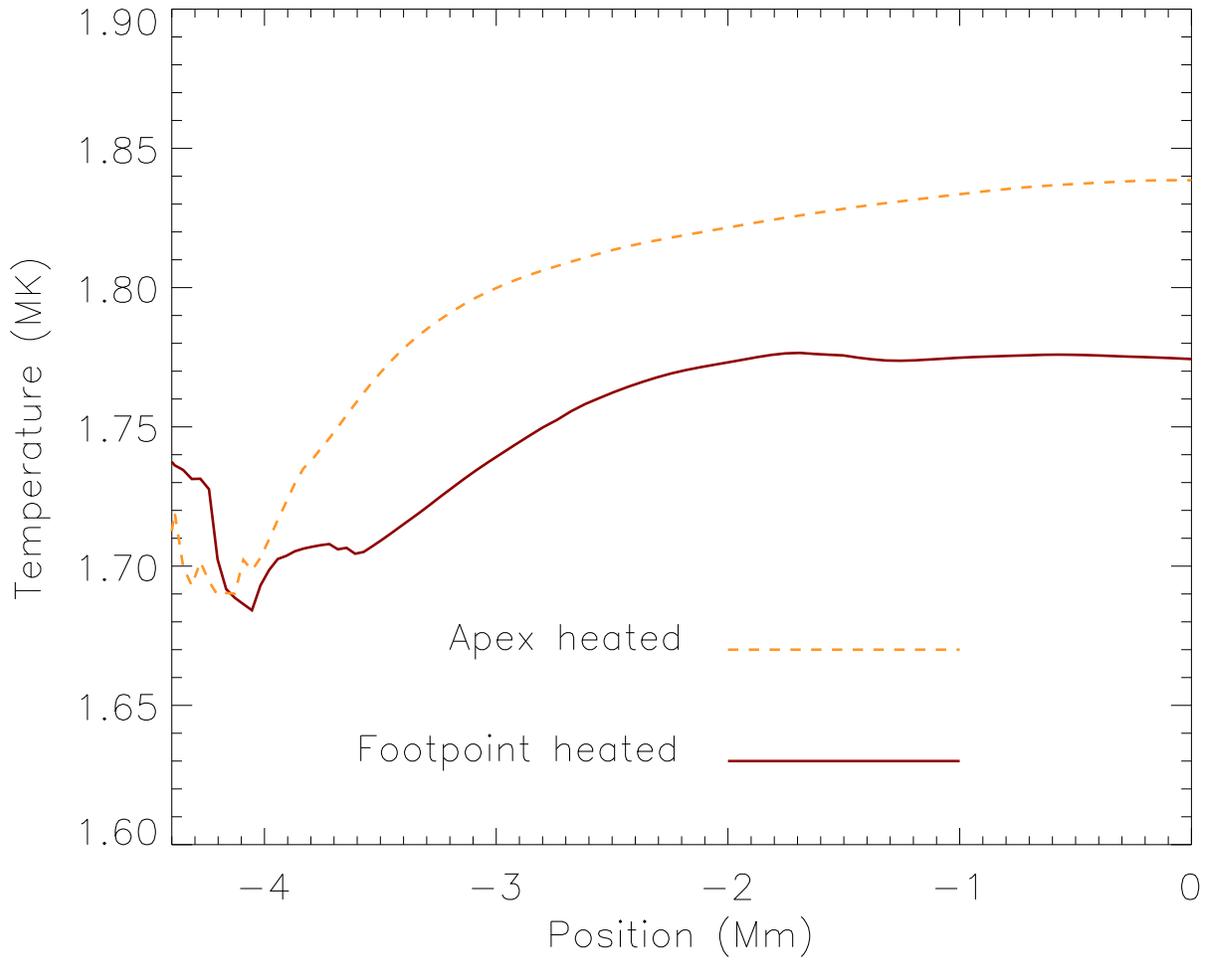}
\caption{Temperature profile along the length of the loop derived using 284/195 filter ratio. \label{fig5}}
\end{figure}

Considering each in turn, we see firstly that the single filter ratio technique gives three quite different temperature values from the three ratios (Figure \ref{fig4}). It must be noted that this has been investigated by \citet{weber05} who employed a flat differential emission measure distribution to mimic an ``extreme'' multithermal plasma. The paper found clearly that the single filter ratio method is very biased towards the ratio of the integrals of the temperature response function under these conditions.
Essentially the same scenario is happening in the current simulations. Multiple strands are heating up and cooling down simultaneously such that the spread of the true temperature and density values generated from all the strands will be quite broad. Although it will not be completely flat, the resulting DEM stretches over a several decades of temperature (i.e. $\sim 10^4-10^7$K); thus the single filter ratio method will tend towards isothermal values for each given line pair.

\begin{figure}
\centering
\includegraphics[width=0.65\textwidth,angle=90]{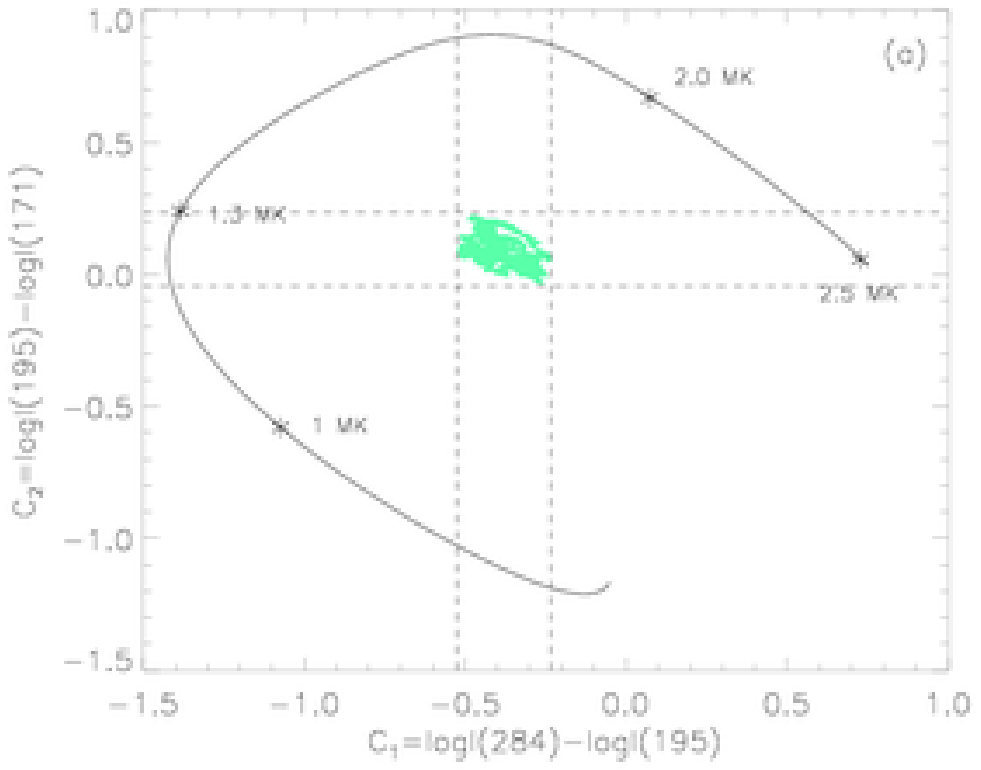}
\includegraphics[width=0.65\textwidth,angle=90]{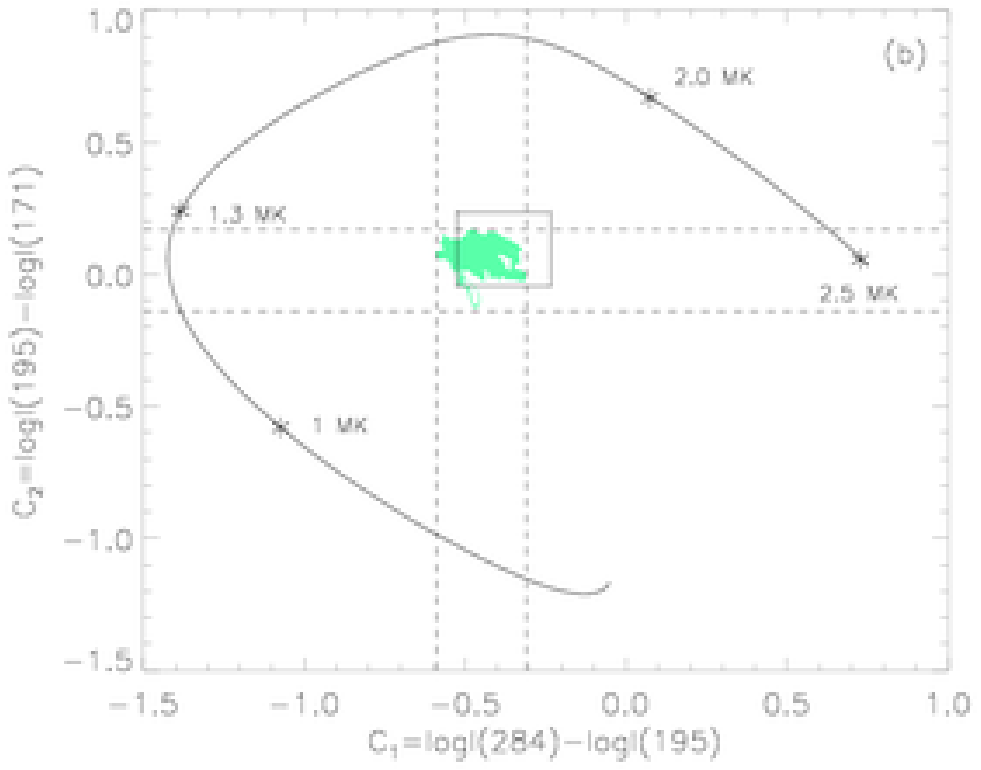}
\caption{Simulated data points in green on the C-C plot for (a) ADH and (b) FDH loop. Note that the solid line is the C-C temperature curve. \label{fig6}}
\end{figure}

\begin{figure}
\centering
\includegraphics[width=0.85\textwidth,angle=90]{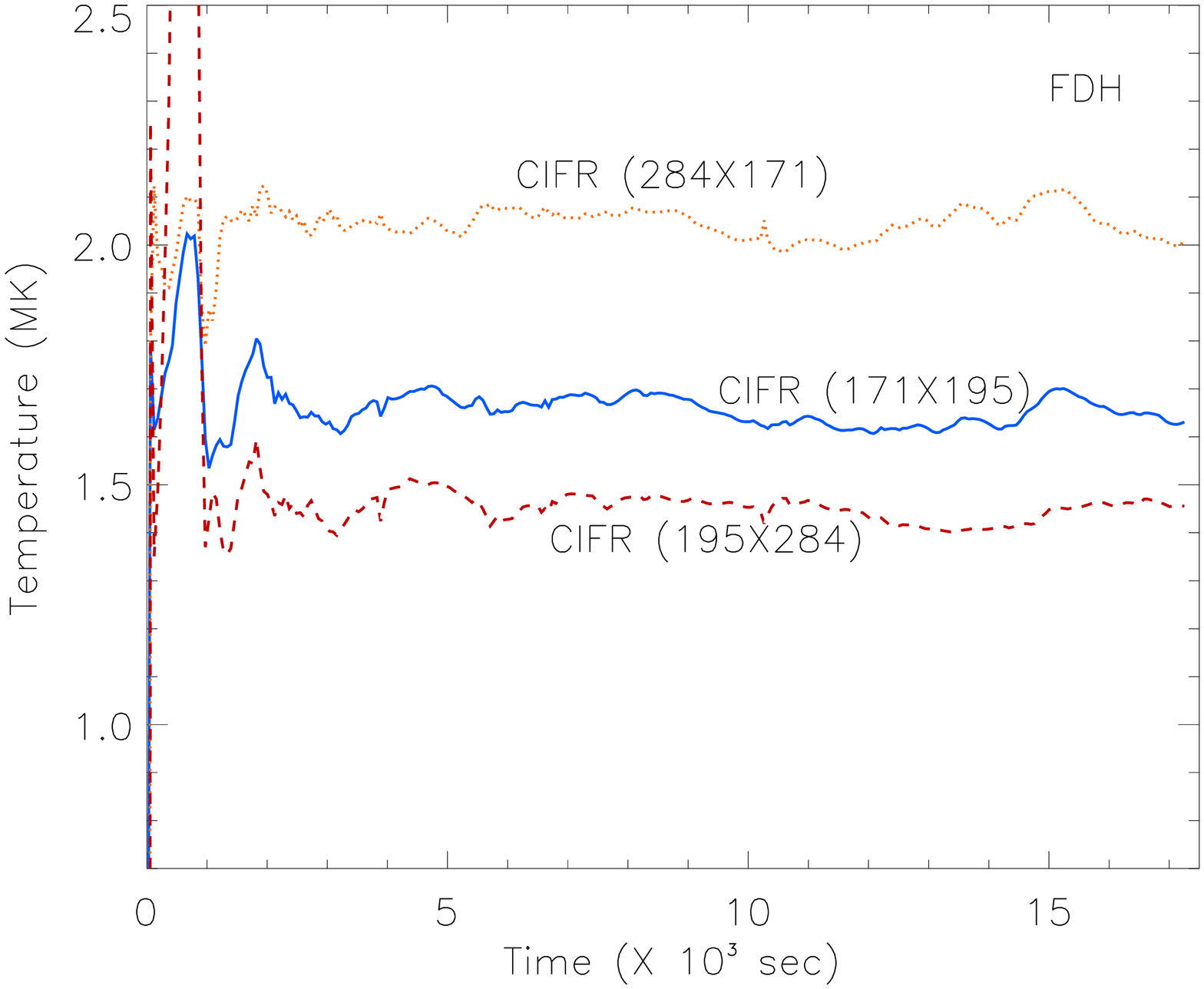}
\caption{Combined improved filter ratio temperature evolution using three TRACE passbands for the FDH case. \label{fig7}}
\end{figure}

Similarly, a complementary argument can be made concerning the double filter ratio method. If there is a tendency towards a certain value of the ratio from a single filter ratio, this will be reflected in a clustering of the points in the C-C curve, as demonstrated in Figure \ref{fig6}. 

Finally consider the CIFR approach. Employing a flat DEM model as a proxy for a multithermal plasma  (as in \citet{weber05}) we can write from \eqref{eq:cfr};
\begin{subequations}
\begin{eqnarray}
 CFR_{171}(T)=\dfrac{(I_{171} \times I_{195} \times I_{284})^{1/3}}{I_{171}} \\
 CFR_{195}(T)=\dfrac{(I_{171} \times I_{195} \times I_{284})^{1/3}}{I_{195}}
\end{eqnarray}
\end{subequations}
Since, 

\begin{equation*}
I_{\lambda} \sim \int G_{\lambda}(T)DEM(T)dT=DEM_{flat}\int G_{\lambda}(T)dT
\end{equation*}

then following equation \eqref{eq:cifr} we will have the CIFR for 171 and 195 as,

\begin{equation}
\label{eq:calccifr}
\begin{split}
CIFR (T)  =\dfrac{(I_{171} \times I_{195} \times I{284})^{2/3}}{I_{171} \times I_{195}} \\
 =\dfrac{(\displaystyle \int_{T_a}^{T_b}G_{171}(T)dT \times \int_{T_a}^{T_b}G_{195}(T)dT \times \int_{T_a}^{T_b}G_{284}(T)dT)^{2/3}}{\displaystyle \int_{T_a}^{T_b}G_{171}(T)dT \times \int_{T_a}^{T_b}G_{195}(T)dT}
\end{split}
\end{equation}
Here, $T_a$ and $T_b$ are the starting temperature and end temperature for the integration. If $T_a$ and $T_b$ are fixed, CIFR($171 \times 195$) always maintains a fixed ratio i.e., $0.6$, which corresponds to the temperature $\sim 1.7$ MK the value we obtain from the simulations. The same is true for all other combination of filters. Thus like the single filter ratio, CIFR also gives the ratio of areas under the temperature response function curves. Hence, one can conclude that for a broad multithermal plasma, CIFR also will provide a flat temperature profile, that does not reflect the real thermal structure.

 In this paper we have produced the CIFR using only three passbands of TRACE. Thus even though our present result is restricted to within the TRACE passbands, the theoretical calculation will be true for Hinode-XRT passbands as well. A detail calculation considering the XRT passbands is in progress for our future work. 

Overall, this analysis demonstrates that the emission and filter ratio results we are getting from imagers can be reconciled with a multi-strand, multi-thermal loop approach. However the temperature values they produce do not correspond at all to the EM weighted temperatures, but are purely an instrumental effect.

\section{Acknowledgments}

This work was supported by the Science and Technology Facilities Council Standard Grant (PP/C502506/1). Authors would like to thank Silvia Dalla and the anonymous referee for their careful reading of the manuscript.

\bibliographystyle{aj}
\bibliography{sw09}

\end{document}